\documentclass{epl}

\title{\bf Potential energy landscape of finite-size mean-field models
              for glasses}

\author{A. Crisanti\inst{1}\footnote{e-mail:andrea.crisanti@phys.uniroma1.it}
        and
        F. Ritort\inst{2}\footnote{e-mail:ritort@ffn.ub.es}}

\institute{
\inst{1} Dipartimento di Fisica, Universit\`a di Roma ``La Sapienza'',
	 and \\
         Istituto Nazionale Fisica della Materia, Unit\`a di Roma I \\
         P.le Aldo Moro 2, I-00185 Roma, Italy.
\inst{2} Physics Department, Faculty of Physics \\
         University of Barcelona, Diagonal 647, 08028 Barcelona,  Spain
        }

\rec{}{}


\begin{document}

\maketitle

\begin{abstract}
We analyze the properties of the energy landscape of finite-size fully
connected spin-glass models with a discontinuous transition.  In the
thermodynamic limit the equilibrium properties in the high temperature
phase are described by the schematic Mode Coupling Theory of
super-cooled liquids.  We show that {\it finite-size} fully
connected spin-glass models do exhibit properties typical of
Lennard-Jones systems when both are near the critical glass transition,
where thermodynamics is ruled by energy minima distribution.  Our
study opens the way to consider activated processes in real glasses
through finite-size corrections (i.e. calculations beyond the saddle
point approximation) in mean-field spin-glass models.
\end{abstract}

In recent years a significant effort has been devoted to the understanding 
of glass-forming systems. Recent theoretical and numerical results
clearly show that the slowing down of the dynamics is strongly connected
to the potential energy landscape. The trajectory of the representative 
point in the configuration space can be viewed as a path in a multidimensional
potential energy surface. The dynamics is therefore strongly influenced by the
topography of the potential energy landscape: local minima, barriers heights,
basin of attraction an other topological properties all influence the 
dynamics. 

The potential energy surface of a super-cooled liquid 
contains a large number of local minima, called {\it inherent structures}
(IS) \cite{S95}, each surrounded by a {\it basin} defined as the set of 
all configurations that a local energy minimization maps onto the IS
contained within it. 
In this picture the time evolution
of the system can be divided into {\it intra-basin} and {\it inter-basin} 
motion. Transition from one basin to another are expected to occur 
differently as the temperature is varied. In particular, when the
temperature is lowered down to the order of the critical Mode Coupling 
Theory (MCT) temperature $T_{MCT}$ the two motions become well separated
in time and the relaxation dynamics is dominated by slow 
thermally activated crossing of potential energy barriers between different
basins \cite{SSDG99}.

The essential features of MCT for glass-forming systems 
are also common to some fully connected spin glass models, 
called mean-field $p$-spin glass with $p>2$ \cite{KT87,CHS93}. 
In the thermodynamic limit the high-temperature phase, 
(paramagnetic in the spin-glass language and liquid in glass language), 
is described by the schematic mode MCT for super-cooled liquids 
\cite{G84,CHS93}. 
As a consequence at the critical temperature, called $T_D$ in $p$-spin 
language and $T_{MCT}$ in MCT language, an ergodic to non-ergodic transition 
takes place. 
In mean-field models the barriers separating different ergodic
componets are proportional to the system size, hence infinite in the
thermodynamic limit, and at $T_D$ the relaxation time diverges.
We note that despite this strong ergodicity beaking the dynamical 
behaviour is far from being trivial \cite{CHS93,CK93}.
In real systems the barriers are of
finite height and the transition to a glassy
state appears at $T_g<T_D$, the glass transition temperature, 
where the typical activation time over barriers is of the same order as 
the observation time. 

Despite these differences mean-field models, having the clear advantage
of being analytically tractable, have been used to study the phase space
structure of glassy systems, especially between the dynamical
temperature $T_D$ and the static temperature $T_c$ (Kauzmann temperature
$T_K$ in glass language) where a real thermodynamic phase transition
driven by the collapse of the configurational entropy takes place.  The
main drawback is that, since activated process cannot be captured by
mean-field models, the picture that emerges is not complete.  Let us
remark also that, despite the large amount of analytical work devoted to
the study of the static as well as dynamical properties in the
$N\to\infty$ limit, much less is known concerning the finite $N$
behavior.

In this Letter we investigate numerically {\it finite-size} fully
connected spin-glass models, where activated process are present.  We
find strong indications that, once activated process are allowed,
mean-field spin glass models with a discontinuous transition
\cite{KT87} do exhibit properties similar to the ones recently
observed in simulations of Lennard-Jones systems near $T_{MCT}$
\cite{SST98,KST99,SKT99,BH99}, making these systems highly valuable
for studying the glass transition not only at MCT level but also in
the activated regime. 

Here we shall focus on the equilibrium properties of the Ising-spin
Random Orthogonal Model (ROM) \cite{MPR94,PP95} which shows a discontinuous
spin-glass transition identical to that found in $p$-spin models
\cite{REV}.  Similar results are, however, obtained using other
$p$-spin-like models, as for example the Bernasconi Model and the Ising
$p$-spin model.  Consequently, our conclusions and results should be
generally valid for mean-field spin glasses with a discontinuous
transition.  The advantage of the ROM lies in its interaction term
which is two-body, at difference with the $p$-body interaction of
$p$-spin models, making the simulations much faster.  Moreover, the ROM
has a very strong freezing glass transition \cite{MPR94} making it an
extremely good microscopic realization of the Random Energy Model
\cite{DERRIDA}.

The model is
defined by the Hamiltonian
\begin{equation}
\label{eq:ham}
  H = - 2 \sum_{ij} J_{ij}\, \sigma_i\, \sigma_j 
\end{equation}
where $\sigma_i=\pm 1$ are $N$ Ising spin variables, 
and $J_{ij}$ is a $N\times N$ random 
symmetric orthogonal matrix with $J_{ii}=0$.  Numerical simulations are
performed using the Monte Carlo method with the Glauber algorithm.
For $N\to\infty$ this model has the
same thermodynamic properties of the $p$-spin model:
a dynamical transition at $T_D=0.536$, 
with threshold energy per spin $e_{th} = E_{th}/N = -1.87$, 
and a static transition at
$T_c=0.25$, with critical energy per spin $e_{1rsb}= -1.936$ 
\cite{MPR94,PP95}. 

The two transitions can be understood from a ``geometrical'' point of
view. The TAP analysis of these models \cite{CS95,PP95} reveals 
that the phase space visited
is composed by an exponentially large (in $N$) number 
of different basins, separated 
by infinitely large (for $N\to\infty$) barriers. Different basins are 
unambiguously labelled by the value of the energy density $e$ at $T=0$. 
Each TAP solution describes the thermodynamics within the basins labelled by 
$e$ and at $T=0$ coincide with the potential energy local minima, i.e. 
the  IS of the system. 
The dynamical transition is associated with IS with the
largest basin of attraction for $N\to\infty$, while the static transition
with IS with the lowest accessible free energy \cite{CS95,KW87}. 

In the mean-field limit, the allowed values of $e$ are between 
$e_{1rsb}$ and $e_{th}$. Solutions with $e$ larger than $e_{th}$ are unstable
\cite{CGP98},
while solutions with $e$ smaller than $e_{1rsb}$ have negligible statistical
weight \cite{CS95,PP95}. 
The IS with energy equal to the threshold energy $e_{th}$ can be 
identified with the less stable solutions. 
These IS have the largest (exponentially with $N$)
statistical weigth so that a rapid cooling from an equilibrium state 
at $T>T_D$ down to $T=0$ will, with probability one for $N\to\infty$,  
drives the system to IS with $e=e_{th}$. 
For finite $N$ the scenario is different since not only IS with $e < e_{th}$ 
acquire statistical weigth, but solutions with $e>e_{th}$ 
and few negative directions \cite{CGP98} may become stable, 
simply because there are not enough degrees of freedom to hit them. 

To analyze the thermodynamics of finite systems we follow the ideas of 
of Stillinger and Weber \cite{SW82} and decompose the
partition sum into a sum over different IS and
a sum within each basin.
Collecting all IS with the same energy
$e$, denoting with $\exp [N s_c(E)]\,de$ the number of IS 
with energy between 
$e$ and $e+de$, and shifting the energy of each basin with
that of the associated IS,
the partition sum can be written as \cite{SW82}
\begin{equation}
\label{eq:part}
  Z_N(T)\simeq \int d e \exp\, N\,\left[-\beta e + s_c(e)
                                     -\beta f(\beta,e)
                               \right]
\end{equation}
where $f(\beta,e)$ can be seen as the free energy density of the
system when confined in one of the basin associated with IS of energy
$e$. The function $s_c(e)$ is the {\it configurational entropy
density} also called {\it complexity} (the same derivation can be also
obtained within the replica approach \cite{CMPV}). From
eq. (\ref{eq:part}) we easily obtain the probability that an
equilibrium configuration at temperature $T=1/\beta$ lies in a basin
associated with IS of energy between $e$ and $e+de$:
\begin{equation}
\label{eq:prob}
 P_N(e,T) =  \exp\, N\,\left[-\beta e + s_c(e)
                                     -\beta f(\beta,e)
                               \right] / Z_N(T).
\end{equation}
Taking the $N\to\infty$ limit we recover the mean-field
results \cite{CS95,MPR94,PP95}. Here we keep $N$
{\it finite}.

\begin{figure}
\onefigure{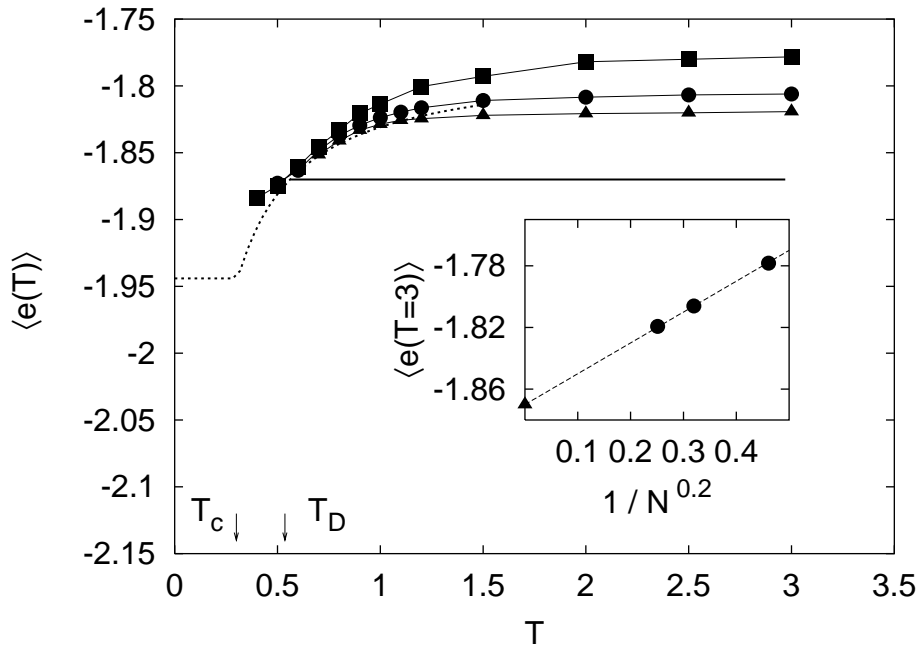}
\caption{Temperature dependence of $\langle e(T)\rangle$ for
         $N=48$ (square), $N=300$ (circle) and $N=1000$ (triangle). 
	The average is over $10^3$ different equilibrium
	configurations at temperature $T$. 
	 The horizontal line is the $N\to\infty$ limit.
         The arrows indicate the critical temperatures $T_D$  
         and $T_c$ (see text). 
         The dotted line is the curve obtained from the
         configurational entropy for large $N$.
         In the inset it is shown the size dependence of 
         $\langle e(T=3)\rangle$ as a function $N^{-\alpha}$ 
         with $\alpha=0.2$ extrapolated down to $-1.87$
         (triangle).
}
\label{fig:f1}
\end{figure}

  From the partition function we can compute the average
internal energy density. This is given by 
$u(T) = \langle e +\partial (\beta f) / \partial \beta \rangle =$
$\langle e(T) \rangle + \langle \Delta e(T)\rangle$, where the first term is 
the average energy of the IS relevant for the thermodynamics 
and the second is the contribution coming out from fluctuations within the IS.
The average is taken with the weight 
(\ref{eq:prob}). 
In the limit $N\to\infty$ the only relevant IS are those with $e=e_{th}$,
and $\lim_{N\to\infty} \langle e(T) \rangle = e_{th}$ for any $T>T_D$. 
To measure $\langle e(T) \rangle$ for finite $N$ we consider the following 
experiment. First we equilibrate the system at a given temperature $T$, then we
instantaneously quench it at $T=0$. This is obtained by decreasing the
energy along the steepest descent path. In this way we can identify the
energy of the IS. The experiment is repeated several times starting from
uncorrelated equilibrium configurations at $T$ and the average IS energy
is computed. 
In figure \ref{fig:f1} we report $\langle e(T)\rangle$ as a function 
of temperature $T$ for system sizes $N=48$, $300$ and $1000$. 
As expected, as $N$ increases $\langle e(T)\rangle$
tends to $e_{th}$. From the numerical data we found that the plateau energy,
approaches $e_{th}$ with the power law
$\langle e_{plateau}\rangle - e_{th} \sim N^{-0.2}$, see inset 
Fig. \ref{fig:f1}.
Note that 
since $N$ is finite we can equilibrate the system also 
below $T_D$, down to the glassy transition $T_g(N)$, $\sim 0.35$ for $N=48$ 
and $\sim 0.5$ for $N=300$, 
below which the system falls out of equilibrium \cite{Note1}.

The figure shows that for finite $N$ and $T$ not too close to
$T_D$ the thermodynamics is dominated by IS with $e>e_{th}$. This is 
clearly seen from the (equilibrium) probability distribution of $e$
since it is centered about $\langle e(T)\rangle$ indicating that IS
with $e\simeq \langle e(T)\rangle$ have the largest statistical weight.
This scenario has been also observed in numerical simulation of,
e.g., Lennard-Jones systems \cite{OTW88,SDS98,SST98,KST99}.

\begin{figure}
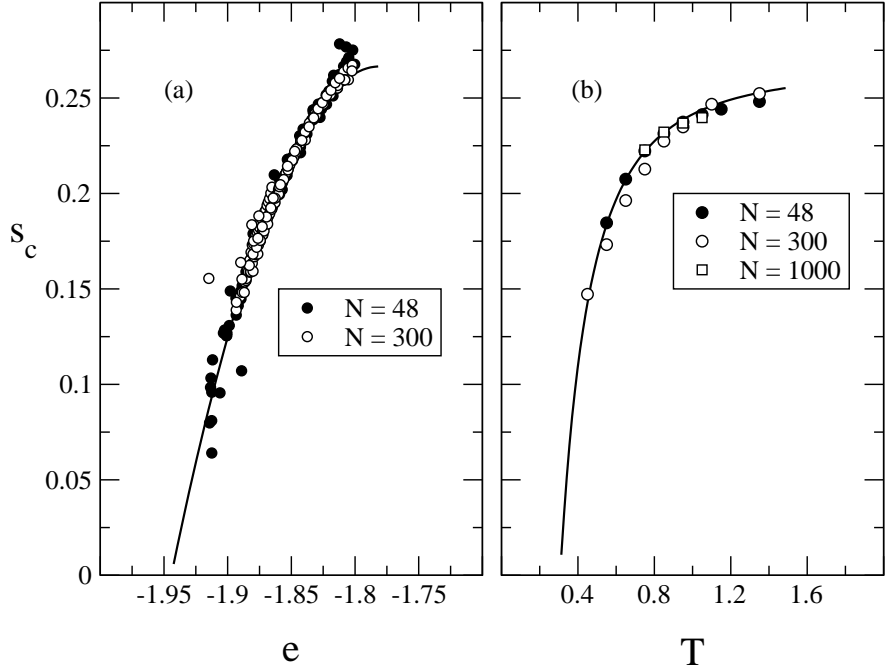

\onefigure{rom_glass_f2.eps}
\caption{(a) Configurational entropy as a function of energy.
         The data are from system sizes $N=48$ and $N=300$,
         and temperatures $T=0.4$, $0.5$, $0.6$, $0.7$, 
         $0.8$, $0.9$ and $1.0$. For each curve the unknown constant has
         been fixed to maximize the overlap between the data and
         the theoretical result \protect\cite{PP95}. The line is the
	 quadratic best-fit.
         (b) Configurational entropy density as a function of temperature.
         The line is the result from the best-fit of $s_c(e)$ 
         while the symbols are the results from the temperature integration
         of eq. (\protect\ref{eq:entro}) for $N=48$,
         $N=300$ and $N=1000$.
}
\label{fig:f2}
\end{figure}

In the temperature range where eq. (\ref{eq:prob}) is valid, the
curves $\ln P_N(e,T) + \beta e$ are equal, except for a temperature
dependent factor $\ln Z_N(T)$, to $s_c(e)-\beta f(\beta,e)$. If we can
neglect the energy dependence of $f(\beta,e)$, then it is possible to
superimpose the curves for different temperature. The resulting curve
is, except for an unknown constant, the complexity $s_c(e)$.  The
curves obtained for system sizes $N=48$ and $300$ and various
temperatures between $T=0.4$ and $T=1.0$ are shown in figure
\ref{fig:f2} (a). The data collapse is rather good for $e <
-1.8$. Above the curves cannot be superimposed indicating that the
energy dependence of $f(\beta,e)$ cannot be neglected anymore.  In
liquid language this is the anharmonic threshold \cite{SKT99,BH99}.
To compare the result with the analytical predictions each curve has
been translated to maximize the overlap with the theoretical
prediction for $s_c(e)$ \cite{PP95}.  The line is the quadratic
best fit from which we can estimate the critical energy $e_c\simeq
-1.944$, where $s_c(e_c)$ vanishes, in good agreement with the
theoretical result $e_{1rsb}=-1.936$ \cite{PP95}.

Direct consequence of $f(\beta,e) \simeq f(\beta)$ for $e<-1.8$ is
that in this range the partition function can be written as the
product of an intra-basin \cite{Note2} contribution [$\exp(-N\beta
f$)] and of a configurational contribution which depends only on the
distribution of IS energy densities.  The system can then be
considered as composed by two independent subsystems: the intra-basin
subsystem describing the equilibrium when confined within basins, and
the IS subsystem describing equilibrium via activated processes
between different basins.  As the temperature is lowered and/or $N$
increased the two processes get more separated in time and the
separation into two subsystems becomes more and more accurate.  A
similar scenario is also oserved in numerical simulations of 
super-cooled liquids,
such as Lennard-Jones systems, near the MCT
transition \cite{ST97,SSDG99}.  
The form of $f(\beta)$ for the specific system can be computed 
studying the motion near the IS, for example
using an harmonic approximation \cite{SKT99,BH99}. However, this usually 
gives only a small corrections to thermodynamic quantities for $T$ close to 
$T_D$ \cite{SKT99,BH99} and we do not consider it in this Letter.

Another important consequence of the separation into two subsystems 
is that in eq. (\ref{eq:prob}) $f(\beta,e)$ can be neglected since 
it cancels with the equal term coming from the denominator.
Therefore from the knowledge of $s_c(e)$ we can easily compute 
the average IS energy density $\langle e(T)\rangle$. 
For large $N$ this is given by the saddle-point estimate, 
see eq. (\ref{eq:prob}):
\begin{equation}
\label{eq:eav}
 \max_{e}\, \left[-\beta\, e + s_c(e) \right].
\end{equation}
The result obtained using for $s_c(e)$ 
the quadratic best-fit,
is the dotted line shown in Fig. \ref{fig:f1}. The agreement with the 
direct numerical data is good already for $N=300$. From the form of 
$\langle e(T)\rangle$ for large $N$ 
we can identify the static critical temperature $T_c\simeq 0.3$
as the temperature below which $\langle e(T)\rangle$ remains constant,
not far from the theoretical results $T_c=0.25$ \cite{PP95}. 
To have a check of our results we have computed the configurational 
entropy density using a different approach \cite{SKT99}. 
When the IS subsystem is in 
thermal equilibrium then the temperature dependence of the configurational
entropy density can be evaluated from the thermodynamic relation
\begin{equation}
\label{eq:entro}
 \frac{d\, s_c(T)}{d\, \langle e(T)\rangle} = \frac{1}{T}
\end{equation}
integrating the $T$ dependence of $d\,\langle e(T)\rangle / T$.
Using the data of figure \ref{fig:f1} we obtain the curves shown in figure
\ref{fig:f2} (b). The line is the result valid for large $N$ obtained from 
the quadratic best fit of $s_c(e)$.
The agreement for $N=300$ and $1000$ is rather good. 
We note that increasing $N$ reduces the IS energy range
explored by the system for a given fixed Montecarlo simulation length, and this
in turn reduces the temperature range in figure \ref{fig:f2} (b). 
In general to have a good resolution of the IS-energy range $N$ should not be
too large, however it cannot be too small to have a good
sampling of the relevant IS. The sizes reported here 
give reasonable results, even if $N=48$ and $N=1000$ are close to the
border. 

\begin{figure}
\onefigure{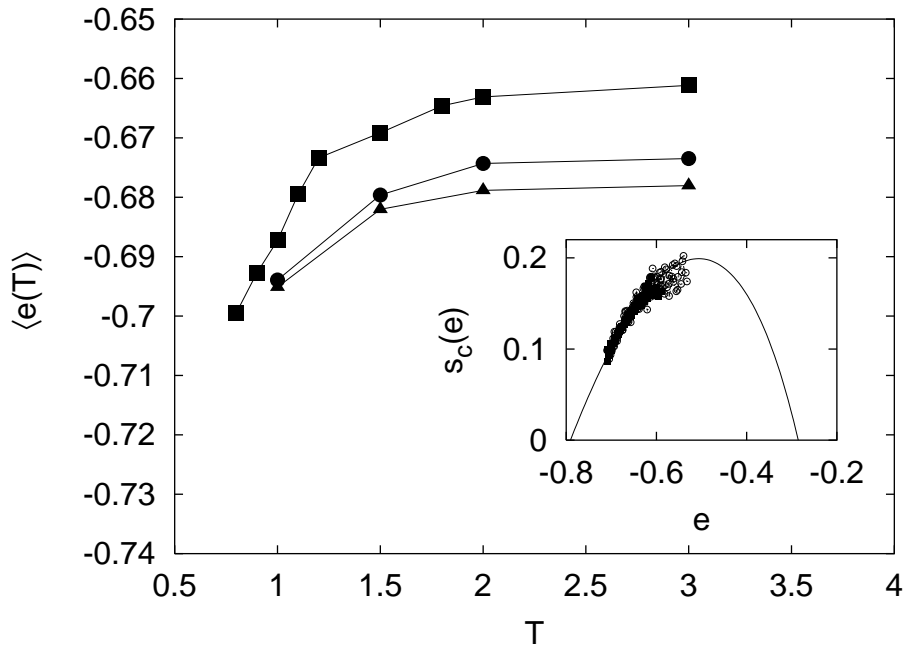}
\caption{Temperature dependence of $\langle e(T)\rangle$ for the
	$\pm J$ SK model and 
         $N=192$ (square), $N=320$ (circle) and $N=640$ (triangle). 
	The average is over $10^3$ different equilibrium
	configuration at temperature $T$. 
        Inset: Configuration entropy as a function of energy for the
         $\pm J$ SK model. 
         The data are from system sizes $N=64$ (circle) and $N=192$
         (square),
	 and temperatures $T=0.8$, $0.9$, $1.0$, $1.1$, $1.2$, 
	 $1.3$. For each curve the unknown constant has
         been fixed to maximize the overlap between the data and
         the theoretical result (continuous line).
         \protect\cite{BM81}.
}
\label{fig:f1sk}
\end{figure}

We finally note that the analysis of the IS done in this Letter can be 
also useful for systems with different types of transition, e.g.
the spin glass transition.  In figure
\ref{fig:f1sk} it is shown the average IS energy density 
$\langle e(T)\rangle$ as function of $T$ for different system sizes 
for the mean-field $\pm J$ Sherrington-Kirkpatrick model,
and in the inset the configurational entropy density $s_c(e)$.
For a more detailed IS-analysis for this model see Ref.\cite{CMPR00}.

To summarize, in this Letter,  
{\bf using an IS-based analysis,} 
we have shown that {\it finite-size}
mean-field spin-glass models with discontinuous transition are good
candidates for studying the glass transition. 
{\bf We note that although the IS approach may be generally
applied to any systems its relevance for thermodynamics is
``a priori'' not obvious. For example simple one-dimensional models with 
non trivial configurational entropy may show uninteresting
thermodynamical behaviour \cite{BIMO}. Moreover, 
Stillinger \cite{STI} has shown that
for realistic systems an IS-based analysis would lead to 
a zero temperature Kauzmann
transition. Both problems seem to be absent for the finite size mean-field 
spin glass models discussed in this Letter.}

Our study opens
the way for addressing activated processes in glasses through
finite-size corrections, i.e., calculations beyond the saddle-point
approximation, in mean-field spin-glass models. Finite-size mean-field
spin glass models, for instance $p$-spin models in the spherical
approximation, have the double advantage of being analytically tractable
for $N\to\infty$ and easily simulated numerically for finite $N$,
offering a good model to analyze the glass transition.

\acknowledgments
We thank for useful discussions and critical reading of the manuscript
C. Donati, U. Marini Bettolo and F. Sciortino. F.R has been supported
by the Spanish Government through the project PB97-0971.



%

\end{document}